# A Paired Phase and Magnitude Reconstruction for Advanced Diffusion-Weighted Imaging

Chen Qian, Zi Wang, Xinlin Zhang, Boxuan Shi, Boyu Jiang, Ran Tao, Jing Li, Yuwei Ge, Taishan Kang, Jianzhong Lin, Di Guo, and Xiaobo Qu*

*Abstract*—**Objective**: Multi-shot interleaved echo planer imaging can obtain diffusion-weighted images (DWI) with high spatial resolution and low distortion, but suffers from ghost artifacts introduced by phase variations between shots. In this work, we aim at solving the challenging reconstructions under inter-shot motions between shots and a low signal-to-noise ratio. **Methods**: An explicit phase model with paired phase and magnitude priors is proposed to regularize the reconstruction (PAIR). The former prior is derived from the smoothness of the shot phase and enforced with low-rankness in the k-space domain. The latter explores similar edges among multi-b-value and multi-direction DWI with weighted total variation in the image domain. **Results**: Extensive simulation and *in vivo* results show that PAIR can remove ghost artifacts very well under a high number of shots (8 shots) and significantly suppress the noise under the ultra-high b-value (4000 s/mm$^2$). **Conclusion**: The explicit phase model PAIR with complementary priors has a good performance on challenging reconstructions under inter-shot motions between shots and a low signal-to-noise ratio. **Significance**: PAIR has great potential in advanced clinical DWI applications and brain function research.

## I. INTRODUCTION

DIFFUSION-WEIGHTED image (DWI) is a non-invasive tool for imaging water molecules diffusion [1]. It has been widely employed in the clinical diagnosis of acute stroke [2, 3] and cancer [4-6] and the scientific research of brain fiber tractography [7, 8]. To achieve high spatial resolution and low distortion DWI, multi-shot interleaved echo planer imaging has become increasingly popular [6, 9]. This imaging scheme samples different segments of k-space uniformly along phase encoding direction in different shots. However, during the data acquisition of each shot, subject or physiological motions are easily introduced, leading to strong phase variations of each shot image and finally producing ghosting image artifacts [10]. These artifacts will be more severe, when the phase variations become much heavier, due to the larger diffusion gradient or longer diffusion time, in the higher b-value DWI [11]. In clinical research, lesions were observed more conspicuous in higher b-value DWI [3]. Therefore, achieving high spatial resolution and low distortion DWI is very important.

Phase variations can be corrected by navigator-based [10, 12-14] and navigator-free methods [15-22]. The former acquires an extra navigator echo and assumes it has the same phase as the target image. With this phase information, ghosting artifacts in the target image can be eliminated in the reconstruction [10]. However, between the navigator echo and target image, there commonly exists geometric mismatches that need to be compensated [12, 14]. Also, navigator echo suffers from a low signal-to-noise ratio (SNR) as the signal intensity decreases exponentially when the b-value increases. Thus, eliminating ghosting artifacts and suppressing noise are challenging under high b-value DWI.

Navigator-free methods have been attached with increasing attention recently [15-22]. They can be roughly classified into DWI image reconstructions with the implicit [15-18] or explicit phase [19-22] (Fig. 1).

Implicit phase reconstructions recover the image of each shot and then combine them into a magnitude image by the sum of squares (SOS) [15-18]. These approaches avoid the estimation of the phase of each shot image, i.e., the shot phase. To reconstruct each shot image, parallel imaging was employed at the early stage. But the shot number was restricted to two or three [15] and reconstructed images were still in high distortion and low spatial resolution. Recently, inspired by models and priors in fast magnetic resonance imaging [23-26], many state-of-the-art methods exploit low-rank properties in multi-shot interleaved echo planer imaging DWI. MUSSELS assumes that phase modulation between shots is smooth and approximates this modulation with a limited support convolution kernel in k-space [16]. This convolution indicates an annihilation relationship that is deduced for the structured low-rank matrix lifting. Then, each shot image is reconstructed by filling missing entries using low-rank matrix completion approaches [16]. PLRHM builds a Hankel matrix based on the assumption

---

This work was supported in part by the National Natural Science Foundation of China under grants, 61971361, 62122064, and 61871341, the Natural Science Foundation of Fujian Province of China under grants 2021J011184, the President Fund of Xiamen University(0621ZK1035), and the Xiamen University Nanqiang Outstanding Talents Program.

Chen Qian, Zi Wang, Xinlin Zhang, Boxuan Shi, and Xiaobo Qu* are with the Department of Electronic Science, Biomedical Intelligent Cloud R&D Center, Fujian Provincial Key Laboratory of Plasma and Magnetic Resonance, National Institute for Data Science in Health and Medicine, Xiamen University, China (*Corresponding author, email: quxiaobo@xmu.edu.cn).

Boyu Jiang and Ran Tao are with the United Imaging Healthcare, Shanghai, China.

Jing Li and Yuwei Ge are with Xingaoyi Medical Equipment Co., Ltd., Yuyao, China.

Taishan Kang and Jianzhong Lin are with Department of Radiology, Zhongshan Hospital of Xiamen University, School of Medicine, Xiamen University, Xiamen, China.

Di Guo is with the School of Computer and Information Engineering, Xiamen University of Technology, Xiamen, China.

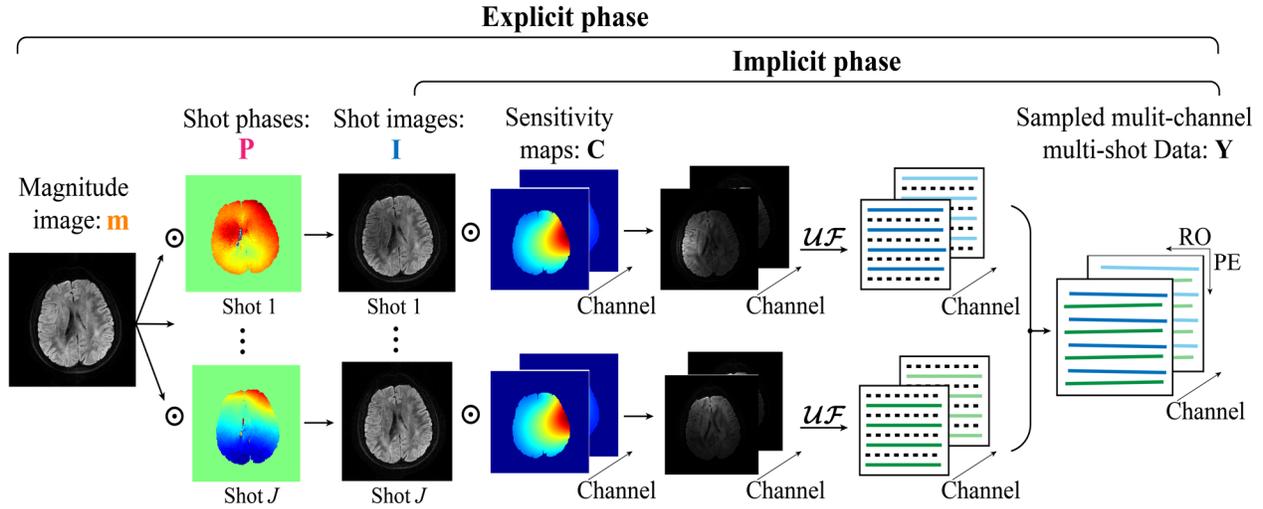

Fig. 1. Illustration of explicit and implicit shot phase reconstruction strategy. Note: RO is readout and PE is phase encoding, $\mathcal{U}$ is an undersampling operator that fills zeros on non-acquired data points, and $\mathcal{F}$ is a Fourier transform operator.

that the phase of magnetic resonance images is smooth [18]. Then, it constrains the low-rankness of this matrix for each shot image by minimizing partial sum minimization of singular values [18].

Explicit phase reconstructions estimate the shot phase first and then incorporate all shot data to build an integrated reconstruction problem. Finally, a magnitude image is estimated assuming that different shot images share the same magnitude [19-21]. A representative method is MUSE which reconstructs each shot image by SENSE [27] (in-plane under-sampling factor equals the shot number) and denoises the shot phase with a total variation [19]. Then, the shot phase is employed to incorporate all shot data into a unified reconstruction problem for the magnitude image.

Compared with implicit phase reconstruction, the explicit phase strategy decreases the number of unknowns, which improves matrix inversion conditioning [19-21]. This strategy may bring benefits in low SNR imaging scenarios, such as ultra-high b-value DWI.

An accurate phase estimation is important for explicit phase reconstruction. The step-by-step estimation of shot phase and magnitude in MUSE can hardly get reliable shot phases when the shot number is high, say eight [16]. Recently, many methods attempt to iteratively solve the shot phase and magnitude image. POCS-ICE alternatively updates the shot phase and magnitude images in an iterative process [21]. PR-SENSE introduces a total variation regularization to solve the integrated equation [22]. However, when the number of shots or the b-values increases, it is still challenging to obtain reliable shot phases (See results in Section IV.C).

In this work, we aim at reconstructing the shot phase and magnitude image under inter-shot motions and a low signal-to-noise ratio. A model with paired phase and magnitude priors is proposed to regularize the reconstruction (PAIR). The former prior is derived from the smoothness of the shot phase and enforced with low-rankness in the k-space domain. The latter explores similar edges among multi-b-value and multi-direction

DWI with weighted total variation in the image domain.

Our main novelties and contributions are summarized as follows:

1) A explicit phase model PAIR is proposed to exploit two complementary constraints of phase and magnitude (low-rank priors in the k-space domain and weighted total variation in the image domain).

2) Extensive *in vivo* results show that PAIR can remove ghost artifacts very well under a high number of shots (8 shots); suppress the noise significantly under the ultra-high b-value (4000 s/mm$^2$); obtain high-fidelity reconstruction under both uniform and partial Fourier undersampling; achieve nice robustness on multi-vendor multi-center data.

## II. RELATED WORKS

### A. Implicit phase reconstructions

Implicit phase methods reconstruct all shot images without shot phase estimation. They treat the recovery of shot images as an under-sampling reconstruction problem. The forward model is (Fig. 1):

$$\mathbf{Y}_h = \mathcal{U}\mathcal{F}\mathbf{C}_h\mathbf{I} + \boldsymbol{\eta}, \qquad (1)$$

where $\mathbf{Y}_h \in \left[\mathbf{Y}_{h,1}, \dots, \mathbf{Y}_{h,j} \dots \mathbf{Y}_{h,J}\right] \in \mathbb{C}^{MN*J}$ denotes the sampled k-space data of the $h$-$th$ channel, $\mathbf{C}_h \in \mathbb{C}^{MN*MN}$ is the $h$-$th$ coil sensitivity map, $\mathbf{I} = \left[\mathbf{I}_1, \dots, \mathbf{I}_j, \dots \mathbf{I}_J\right] \in \mathbb{C}^{MN*J}$ is the concatenated matrices of the vectorized target shot images, $\boldsymbol{\eta}$ is the measurement noise, $M$ and $N$ are the columns and rows of k-space, $H$ and $J$ are the numbers of channels and shots. $\mathcal{U}$ is an under-sampling operator that fills zeros on non-acquired data points, $\mathcal{F}$ is a Fourier transform operator.

Implicit phase reconstruction models exploit low-rankness properties in multi-shot DWI data [16-18]:

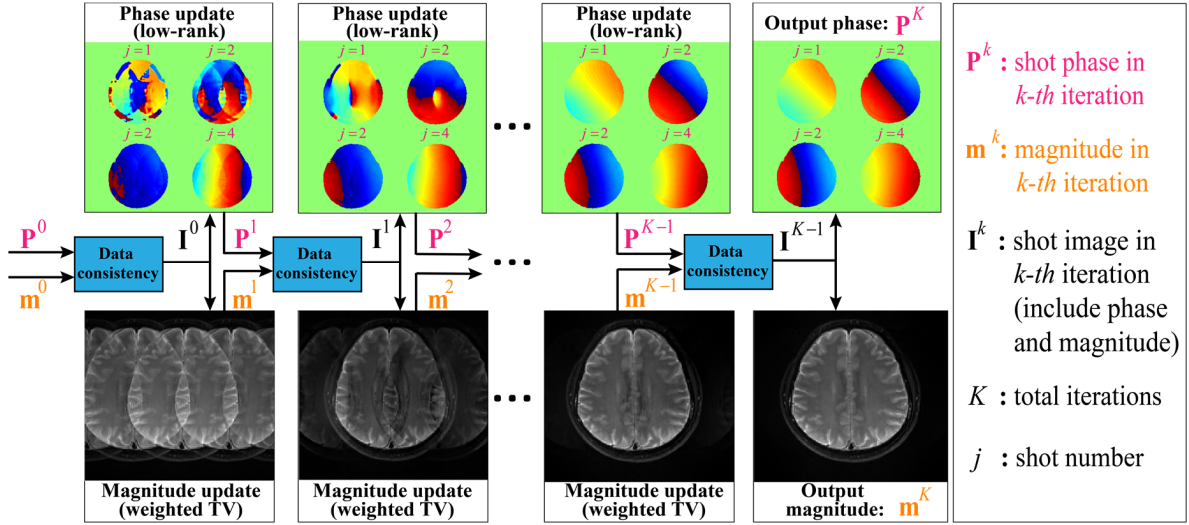

**Fig. 2.** Solving process of the PAIR. Low-rank and weighted total variation are used as a pair of complementary priors to facilitate phase and magnitude image reconstruction. Note: TV is short for total variation.

$$\min_{\mathbf{I}} \lambda \sum_h^H \left\| \mathbf{Y}_h - \mathcal{UFC}_h \mathbf{I} \right\|_F^2 + \mathcal{R}_l \left( \mathcal{FI} \right), \qquad (2)$$

where $\lambda$ balances the data consistency and low-rankness constraint, $\|\cdot\|_F$ represents the Frobenius norm and $\mathcal{R}_l$ is the low-rank regularization.

### B. Explicit phase reconstructions

Explicit phase reconstructions try to get a magnitude image from all shots data, assuming that different shot images share the same magnitude [19-22]. The forward model is (Fig. 1):

$$\mathbf{Y}_{hj} = \mathcal{UFC}_h \mathbf{P}_j \mathbf{m} + \boldsymbol{\eta}, \qquad (3)$$

where $\mathbf{Y}_{hj} \in \mathbb{C}^{MN}$ denotes the sampled k-space data in the vectorized $h$-th channel and $j$-th shot, $\mathbf{P}_j \in \mathbb{C}^{MN*MN}$ is the phase of $j$-th shot, $\mathbf{m} \in \mathbb{R}^{MN}$ is the vectorized target magnitude image. The explicit phase reconstruction model is [19, 21]:

$$\min_{\mathbf{P},\mathbf{m}} \lambda \sum_{h=1}^H \sum_{i=1}^I \left\| \mathbf{Y}_{hj} - \mathcal{UFC}_h \mathbf{P}_j \mathbf{m} \right\|_F^2 + \mathcal{R}_s \left( \mathbf{m} \right), \qquad (4)$$

where shot phase $\mathbf{P}$ and magnitude $\mathbf{m}$ could be updated iteratively. $\mathcal{R}_s$ is the sparse regularization, such as the total variation [22, 28].

In the solving of Eq. (4), when $\mathbf{P}$ is explicitly known, the only variable to solve for is $\mathbf{m} \in \mathbb{R}^{MN}$. Compared with $\mathbf{I} \in \mathbb{C}^{MN*J}$ (Eq. (2)), the number of unknowns in $\mathbf{m} \in \mathbb{R}^{MN}$ (Eq. (4)) is reduced by $J$ times. Thus, an obvious advantage of explicit phase reconstructions is using $\mathbf{P}$ to incorporate all shots data in an integrated equation to improve matrix inversion conditioning, which will bring SNR benefits [19].

The accurate estimation of $\mathbf{P}$ is significant for reconstructing $\mathbf{m}$. However, very few regularizations for $\mathbf{P}$ have been studied in the explicit phase models to the best of our knowledge.

## III. PROPOSED METHOD

In this work, an explicit model with paired shot phase ($\mathbf{P}$) and magnitude ($\mathbf{m}$) priors is proposed to regularize the reconstruction. The whole process is summarized in Fig. 2.

### A. Smoothness of shot phase and induced low-rankness

The shot phases consist of the motion-induced phase and the motion-independent background phase.

The motion-induced phase in the measured multi-shot DWI signal is analyzed first. Positions in the subject and magnetic resonance gradients coordinate systems are defined as $r$ and $R$, respectively [10]. The changes in the relative position between two coordinate systems at spatial location $r$ and time $t$ denotes $R = r + R_0(r,t)$. The measured k-space signal is:

$$y(t) = \int d\mathbf{r} \boldsymbol{\rho}(\mathbf{r}) e^{-i\mathbf{k}(t)\cdot R_0(r,t)} e^{-i\mathbf{k}(t)\cdot r}, \qquad (5)$$

where $\boldsymbol{\rho}(\mathbf{r})$ is relaxation and diffusion-weighted transverse magnetization at $r$, $\mathbf{k}(t)$ is the position in k-space at time $t$.

The accumulated motion-induced phase of $j$-th shot is:

$$\boldsymbol{\phi}_j(\mathbf{r}) = \int R_0(r,t)\cdot\mathbf{k}(t)\,dt = \gamma \int R_0(r,t)\cdot\mathbf{g}(t)\,dt, \qquad (6)$$

where $\gamma$ is the gyromagnetic ratio and $\mathbf{g}$ is the diffusion gradient. The order of $\boldsymbol{\phi}_j$ is only dictated by the variable $r$ in $R_0(r,t)$. When no motion happens ($R_0(r,t) = 0, R = r$), no extra phase is accumulated. When the rigid motion happens, a phase in zero and first order is introduced by shift and rotation, respectively. When more complex motions happen, a higher-order phase will be generated.

To simulate the $\boldsymbol{\phi}$, we employ a second polynomial (Fig. 3(a)), a widely used simulation of the shot phase in multi-shot DWI [18, 22]:

$$\boldsymbol{\phi}_j(x,y) = a_1 + a_2 x + a_3 y + a_4 x^2 + a_5 y^2 + a_6 xy, \qquad (7)$$

where $(x, y) = r$ is the coordinate of the shot phase, $x \in [0, N]$ and $y \in [0, M]$. $(a_1, a_2, a_3, a_4, a_5, a_6)$ is a set of random constant parameters, which are distributed between $[-\pi,\pi)$, $[-\pi/2N,\pi/2N)$, $[-\pi/2M,M)$, $[-\pi/3N^2,\pi/3N^2)$, $[-\pi/3M^2,\pi/3M^2)$, $[-\pi/3NM,\pi/3NM)$, respectively.

The motion-independent background phase $\boldsymbol{\varphi}$ is accumulated by $\rho(r)$ under gradients. Fig. 3(b) shows a background phase $\boldsymbol{\varphi}$ of non-diffusion image $\mathbf{m}_0$ ($b$=0). $\boldsymbol{\varphi}$ is commonly assumed to be smooth and varying slowly [18].

Combing the motion-induced phase $\boldsymbol{\phi}_j$ and motion-independent background phase $\boldsymbol{\varphi}$ to form the shot phase $\mathbf{P}_j$:

$$\mathbf{P}_j = e^{-i\boldsymbol{\phi}_j} \odot e^{-i\boldsymbol{\varphi}} = e^{-i(\boldsymbol{\phi}_j + \boldsymbol{\varphi})}, \tag{8}$$

where $i$ represents imaginary unit. Since both $\boldsymbol{\phi}_j$ and $\boldsymbol{\varphi}$ are commonly assumed to be smooth and varying slowly, $\mathbf{P}_j$ is approximately smooth too. Thus, the $j$-th shot image $\mathbf{I}_j$ can be separated into the shot phase $\mathbf{P}_j$ and the magnitude $\mathbf{m}$ as:

$$\mathbf{I}_j = \mathbf{P}_j\mathbf{m}. \tag{9}$$

The shot phase $\mathbf{P}_j$ is a diagonal matrix, and based on the same magnitude $\mathbf{m}$, we get:

$$\left(\mathbf{P}_j\mathbf{P}_j^*\right)\mathbf{m} = \left(\mathbf{P}_j^*\mathbf{P}_j\right)\mathbf{m}, \tag{10}$$

where superscript * is the complex conjugate.

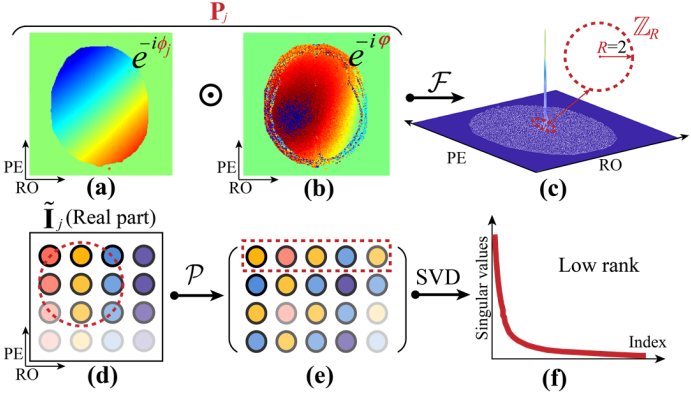

Fig. 3. Low-rankness is deduced from shot phase smoothness. (a) is the motion-induced phase variation $\boldsymbol{\phi}_j$. (b) is the motion-independent background phase $\boldsymbol{\varphi}$. They are multiplied in the Hadamard product to form the shot phase $\mathbf{P}_j$. (c) The nonzero value distribution of $\mathbf{P}_j$ in the k-space is approximately concentrated in limited support $\mathbb{Z}_R$. The k-space of a shot image (d) is lifted to the structured matrix (e), which is low-rank due to the fast decay of singular values (f). Note: RO is readout and PE is phase encoding, $\mathcal{P}$ is the operator converting complex matrix into a low-rank matrix, and SVD is the singular value decomposition.

Substitute shot image $\mathbf{I}_j = \mathbf{P}_j\mathbf{m}$ and the complex conjugate of the shot image $\mathbf{I}_j^* = \left(\mathbf{P}_j\mathbf{m}\right)^* = \mathbf{P}_j^*\mathbf{m}^* = \mathbf{P}_j^*\mathbf{m}$ into Eq. (10), we get:

$$\mathbf{P}_j\mathbf{I}_j^* = \mathbf{P}_j^*\mathbf{I}_j. \tag{11}$$

Transform the left and right of Eq. (11) into k-space:

$$\tilde{\mathbf{P}}_j \otimes \tilde{\mathbf{I}}_j^* = \tilde{\mathbf{P}}_j^* \otimes \tilde{\mathbf{I}}_j, \tag{12}$$

where $\otimes$ is convolution, $\tilde{\mathbf{I}}_j \in \mathbb{C}^{N*M}$ and $\tilde{\mathbf{P}}_j \in \mathbb{C}^{N*M}$ are the Fourier transform of $\mathbf{I}_j$ and $\mathbf{P}_j$.

Rewrite Eq. (12) into a multiplication form:

$$\sum_{(p,q)\in\mathbb{Z}} \tilde{\mathbf{I}}_j^*(-x-p,-y-q)\tilde{\mathbf{P}}_j(p,q)$$
$$- \sum_{(p,q)\in\mathbb{Z}} \tilde{\mathbf{I}}_j(x-p,y-q)\tilde{\mathbf{P}}_j^*(p,q) = 0, \tag{13}$$

where $(x,y)$ is the coordinate of $\tilde{\mathbf{I}}_j$, $x \in [0,N]$ and $y \in [0,M]$. $(p,q) \in \mathbb{Z}$ is the coordinate of $\tilde{\mathbf{P}}_j$ and $\mathbb{Z}$ is the region where $\tilde{\mathbf{P}}_j(p,q)$ is nonzero.

The nonzero value of $\tilde{\mathbf{P}}_j$ in k-space is concentrated in the limited support due to the smoothness of $\mathbf{P}_j$ in the image. Thus, $\mathbb{Z}$ can be approached by a radius $R$ circle region $\mathbb{Z}_R$ (Figs. 3(c) and (d)). With this approximate representation, Eq. (13) holds an annihilation relationship [23]:

$$\sum_{(p,q)\in\mathbb{Z}_R} \tilde{\mathbf{I}}_j^*(-x-p,-y-q)\tilde{\mathbf{P}}_j(p,q)$$
$$- \sum_{(p,q)\in\mathbb{Z}_R} \tilde{\mathbf{I}}_j(x-p,y-q)\tilde{\mathbf{P}}_j^*(p,q) \approx 0, \tag{14}$$

Extract the real and imaginary parts of Eq. (14) and rewrite them into a matrix multiplication form:

$$\mathcal{P}\left(\tilde{\mathbf{I}}_j\right)\begin{bmatrix}\hat{\mathbf{P}}_j^r \\ \hat{\mathbf{P}}_j^i\end{bmatrix} = \begin{bmatrix}\tilde{\mathbf{I}}_j^{r+} - \tilde{\mathbf{I}}_j^{r-} & \tilde{\mathbf{I}}_j^{i-} - \tilde{\mathbf{I}}_j^{i+} \\ \tilde{\mathbf{I}}_j^{i+} - \tilde{\mathbf{I}}_j^{i-} & \tilde{\mathbf{I}}_j^{r-} - \tilde{\mathbf{I}}_j^{r+}\end{bmatrix}\begin{bmatrix}\hat{\mathbf{P}}_j^r \\ \hat{\mathbf{P}}_j^i\end{bmatrix} \approx \mathbf{0}, \tag{15}$$

$$\begin{cases}\tilde{\mathbf{I}}_j^{r+}(e,f) = \tilde{\mathbf{I}}_j^r(x_e - p_f, y_e - q_f), \\ \tilde{\mathbf{I}}_j^{r-}(e,f) = \tilde{\mathbf{I}}_j^r(-x_e - p_f, -y_e - q_f), \\ \tilde{\mathbf{I}}_j^{i+}(e,f) = \tilde{\mathbf{I}}_j^i(x_e - p_f, y_e - q_f), \\ \tilde{\mathbf{I}}_j^{i-}(e,f) = \tilde{\mathbf{I}}_j^i(-x_e - p_f, -y_e - q_f), \\ \hat{\mathbf{P}}_j^r(f,1) = \tilde{\mathbf{P}}_j^r(p_f, q_f), \hat{\mathbf{S}}_j^i(f,1) = \tilde{\mathbf{P}}_j^i(p_f, q_f),\end{cases} \tag{16}$$

where the superscript $r$ and $i$ represent the real and imaginary parts, respectively. $e \in [1,(N-R)*(M-R)]$ and $f \in [1,N_R]$, $N_R$ is the number of elements in $\mathbb{Z}_R$. $\mathcal{P}$ is the operator converting complex matrix into a low-rank matrix (Figs. 3(e) and (f)).

### B. Phase prior based explicit reconstruction (PHASE)

In our previous work, the above low-rankness has been used in an implicit phase reconstruction model PLRHM [18]:

$$(\text{PLRHM}) \min_{\mathbf{I}_j} \frac{\lambda}{2}\sum_{h=1}^{H}\sum_{j=1}^{J}\left\|\mathbf{Y}_{hj} - \mathcal{U}\mathcal{F}\mathbf{C}_h\mathbf{I}_j\right\|_F^2 + \sum_{j=1}^{J}\left\|\mathcal{P}\mathcal{F}\mathbf{I}_j\right\|_*, \tag{17}$$

where $\lambda$ is the regularization parameter and $\|\cdot\|_*$ is the nuclear norm.

Here, we modify PLRHM into an explicit phase model PHASE by separating $\mathbf{I}_j$ into shot phase $\mathbf{P}_j$ and magnitude $\mathbf{m}$:

$$(\text{PHASE}) \min_{\mathbf{P},\mathbf{m}} \frac{\lambda}{2}\sum_{h=1}^{H}\sum_{j=1}^{J}\left\|\mathbf{Y}_{hj} - \mathcal{U}\mathcal{F}\mathbf{C}_h\mathbf{P}_j\mathbf{m}\right\|_F^2 + \sum_{j=1}^{J}\left\|\mathcal{P}\mathcal{F}\mathbf{P}_j\mathbf{m}\right\|_*, \tag{18}$$

The modified PHASE brings two improvements over PLRHM. First, all shot images sharing the same magnitude are utilized as a prior explicitly in PHASE, while shot images in PLRHM have different magnitudes (Fig. 4 (a)). Second, complementary priors can be exploited on shot phase and magnitude separately (see section III.B).

We conducted *in vivo* and simulation reconstructions to explain the advantages of the first improvement: robustness under partially corrupted shot data and nice noise suppression under low SNR scenarios.

For *in vivo* experiment, an 8-shot, b-value 1000 s/mm² DWI data is employed. Different shot images reconstructed by PLRHM are different (Fig. 4(a)). The $8^{th}$ shot data is corrupted by common zipper artifacts [29] (yellow arrows in Fig. 4(a)), leading to the same artifacts residual in the SOS magnitude combined from all shot images (Fig. 4(b)). If the data rejection [30, 31] is employed as a post-processing strategy to exclude the $8^{th}$ shot image before SOS, the artifacts in the magnitude image will be removed (Fig. 4(c)).

The proposed explicit formulation method PHASE can obtain the final magnitude image without SOS. Thanks to the explicit formulation, all shot images strictly share the same magnitude image in the iterations.

Compared with implicit phase methods, PHASE shows good robustness to the partially corrupted data (Fig. R1(c)) without post-processing.

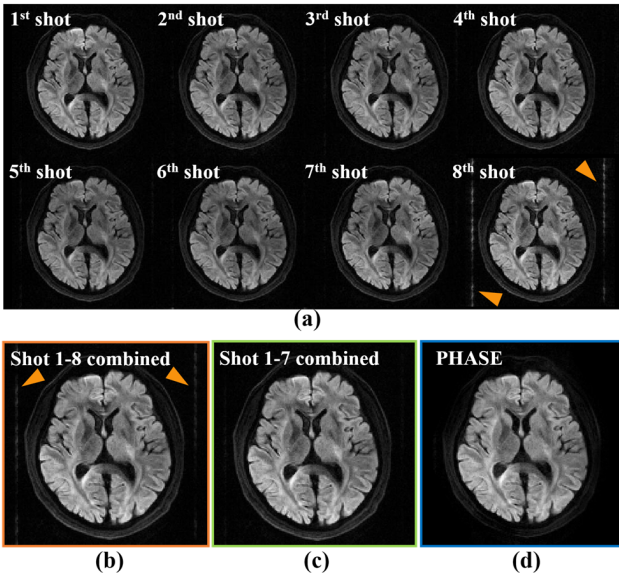

**Fig. 4.** Partially corrupted shot data reconstructions, the data is 8-shot, 17-channel, in-plane resolution 1.0×1.0mm², b-value 1000 s/mm². (a) shows eight shot images reconstructed by PLRHM. (b) and (c) are the SOS magnitude image from all eight shot and first seven shot images, respectively. (c) is the magnitude reconstructed by PHASE.

For simulation comparison, a toy comparison is conducted on a simulated four-shot eight-channel phantom. The whole procedure of simulated multi-shot DWI data is generated as follows: (1) Get a Shepp-Logan phantom image (Fig. 5(a)). (2) Multiply this image with eight-channel coil sensitivity maps (Fig. 5(b)) that are simulated by the Biot-Savart law [22]. (3) Multiply each channel image with the shot phase (Fig. 5(c)) according to Eq. (7). (4) Transform each channel image into its k-space with Fourier transform and then add Gaussian noise to k-space.

PHASE with accurate shot phase ($\mathbf{P}$ is given by simulated shot phase), and PHASE with estimated shot phase are compared with PLRHM [18]. The implicit phase method, PLRHM, loses the image structure and has a large noise residual (Fig. 5(d)). With the proposed PHASE method, this loss is reduced significantly and the noise is suppressed very well (Fig. 5(d)), if the estimated phase is accurate. Even with the estimated phase, the proposed PHASE (Fig. 5(f)) still outperforms PLRHM in lower loss of image structures and better noise suppression. These observations indicate the great potential of explicit phase reconstruction.

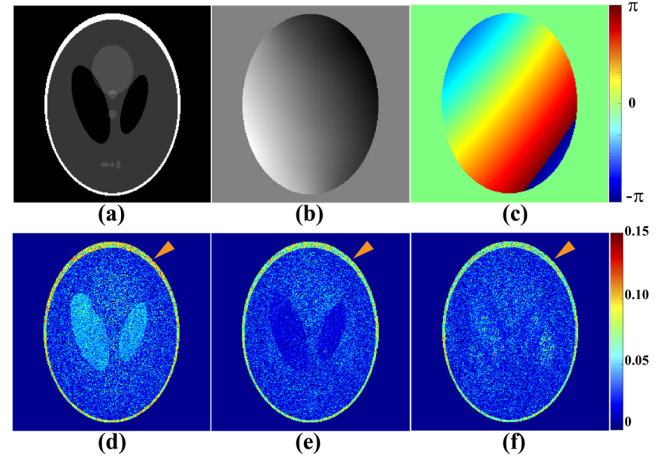

**Fig. 5.** Reconstructions of the simulated four-shot eight-channel phantom. (a) is the ground-truth magnitude image. (b) is one out of eight simulated coil sensitivity maps. (c) is one out of four simulated shot phases. (d)-(f) are error maps of reconstructed images by PLRHM, PHASE with accurate shot phase, and PHASE with estimated shot phase. Note: Gaussian noise (10 dB) is added to the k-space. The PSNRs of (d)-(f) are 31.03, **32.80**, and 32.50, respectively.

### C. Phase And magnitude Image Reconstruction (PAIR)

Besides above phase smoothness prior, the proposed explicit phase model PHASE enables complementary constraint on magnitude $\mathbf{m}$. Thus, we try to combine magnitude prior in the image domain with PHASE to improve the SNR further.

High b-value DWI suffers from low SNR. The signal strength decreases exponentially with the increase of b-value [1]:

$$\boldsymbol{S} = \boldsymbol{S}_0 e^{-b\boldsymbol{D}}, \qquad (19)$$

where $\boldsymbol{S}$ and $\boldsymbol{S}_0$ are voxel signal intensity with and without diffusion-weighted, $b$ is the b-value of $\boldsymbol{S}$, and $\mathbf{D}$ is the diffusion coefficient [1]. Non-diffusion image $\mathbf{m}_0$ ($b$=0) has the highest SNR.

A widely employed magnitude prior is total variation [22, 28]. It can reduce noise by restraining the sharp jump of the input signal [32, 33]. However, total variation may also bring edge blurring to some extent (Figs. 5(a) and (b)).

To avoid edge blur and suppress noise simultaneously, a weighted total variation is added to PHASE to get the paired phase and magnitude image reconstruction (PAIR):

$$\text{(PAIR)} \min_{\mathbf{P},\mathbf{m}} \frac{\lambda}{2} \sum_{h=1}^{H} \sum_{j=1}^{J} \left\| \mathbf{Y}_{hj} - \mathcal{U}\mathcal{F}\mathbf{C}_h \mathbf{P}_j \mathbf{m} \right\|_{\mathrm{F}}^{2} \tag{20}$$

$$+ \sum_{j=1}^{J} \left\| \mathcal{P}\mathcal{F}\mathbf{P}_j \mathbf{m} \right\|_{*} + \beta \left\| \mathbf{m} \right\|_{\mathrm{wTV}} ,$$

$$\left\| \mathbf{m} \right\|_{\mathrm{wTV}} = \sum_{x,y} \left( \mathbf{W}_{\perp}(x,y) \big( \mathbf{m}(x,y) - \mathbf{m}(x-1,y) \big)^2 \right.$$

$$\left. + \mathbf{W}_{=}(x,y) \big( \mathbf{m}(x,y) - \mathbf{m}(x,y-1) \big)^2 \right)^{1/2} , \tag{21}$$

where $\beta$ controls image smoothness. $\mathbf{W}_{\perp}$ and $\mathbf{W}_{=}$ are the weights in vertical and horizontal directions and they represent persistent edges among multi-b-value and multi-direction DWI [34]. In general, $\mathbf{W}_{\perp}$ and $\mathbf{W}_{=}$ are large in the smooth regions and small in the edges.

The prior weights $\mathbf{W}_{\perp}$ and $\mathbf{W}_{=}$ can be calculated from $\mathbf{m_0}$ which has the highest SNR. Moreover, $\mathbf{m_0}$ is usually obtained in the clinic process and has few motion artifacts. The weights are extracted as Eq. (22):

$$\begin{cases} \mathbf{W}_{\perp}(x,y) = \exp\left[ -\big[ \mathbf{m_0}(x,y) - \mathbf{m_0}(x-1,y) \big]^2 \big/ \delta \right] \\ \mathbf{W}_{=}(x,y) = \exp\left[ -\big[ \mathbf{m_0}(x,y) - \mathbf{m_0}(x,y-1) \big]^2 \big/ \delta \right] \end{cases} , \tag{22}$$

where $\delta$ controls weights deviation.

To show the benefits of the edge preserving prior, the methods including PHASE, PAIR with total variation, and PAIR with weighted total variation, are compared in Fig. 5.

PAIR with total variation could greatly suppress the noise but results in loss of edge intensities (white arrow in Fig. 5(b)). PAIR with weighted total variation shows good tolerance to noise and preserves edges much better (Fig.5 (c)). In the following sections, PAIR with weighted total variation is abbreviated as PAIR for short.

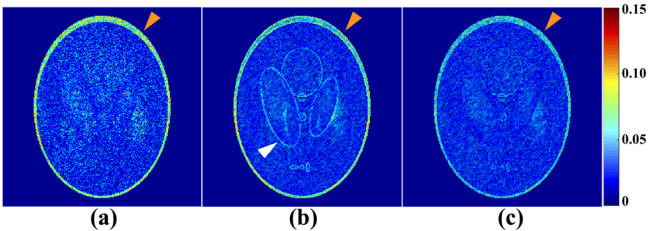

Fig. 6. Reconstructions of the simulated four-shot eight-channel phantom. (a)-(c) are the error maps of PHASE, PAIR with total variation, and PAIR with weighted total variation, respectively. The yellow arrows show different noise residual levels and the white arrow shows blurred edges. Note: Simulated phantom is the same as Fig.4. PSNRs of (a)-(c) are 32.50, 33.12, and **34.23**, respectively.

### D. Numerical algorithm

In this section, we adopt the Projections Onto Convex Sets (POCS) algorithm to solve the PAIR in Eq. (20) [21, 35-37].

The iterative reconstruction is consisting of data consistency, phase update with low-rankness constraint, and magnitude update with weighted total variation (Fig. 2). In the iterative process, the phases become increasingly smooth (upper row in Fig. 2) and artifacts in magnitude image gradually decrease (lower row in Fig. 2).

The k-th ( $k = 1, 2, \ldots K$ ) iteration is shown as follows:

1) Data consistency

$$\mathbf{G}_{hj}^{k} = \mathbf{C}_h \mathbf{P}_j^{k} \mathbf{m}^{k} + \lambda \mathcal{F}^{*} \mathcal{U}^{*} \left( \mathbf{Y}_{hj} - \mathcal{U}\mathcal{F}\mathbf{C}_h \mathbf{P}_j^{k} \mathbf{m}^{k} \right), \tag{23}$$

$$\mathbf{I}_{j}^{k} = \sum_{h=1}^{H} \mathbf{C}_{h}^{*} \mathbf{G}_{hj}^{k}, j = 1, 2, \ldots, J. \tag{24}$$

2) Phase update with low-rankness constraint

$$\mathbf{I}_{j}^{k+1} = \mathcal{F}^{*} \mathcal{P}^{*} \left( \mathrm{SVT}_{\varepsilon} \left( \mathcal{P}\mathcal{F} \left( \mathbf{I}_{j}^{k} \right), \sigma \right) \right), j = 1, 2, \ldots, J, \tag{25}$$

$$\mathbf{P}_{j}^{k+1} = \mathbf{I}_{j}^{k+1} \big/ \left| \mathbf{I}_{j}^{k+1} \right|, j = 1, 2, \ldots, J, \tag{26}$$

where $\mathrm{SVT}_{\varepsilon}(\mathbf{Z}, \sigma)$ is the singular value thresholding operator on a matrix $\mathbf{Z}$ [38, 39]. The first $\varepsilon$ singular values are saved and others minus a proper threshold $\sigma$ .

3) Magnitude update with weighted total variation

$$\mathbf{m}_{avg}^{k} = \sum_{j=1}^{J} \mathbf{P}_{j}^{k*} \odot \mathbf{I}_{j}^{k}, \tag{27}$$

$$\mathbf{m}_{wTV}^{k} = \mathbf{m}_{avg}^{k} - \beta \left( \frac{\partial \left\| \mathbf{m}^{k} \right\|_{\mathrm{wTV}}}{\partial \mathbf{m}^{k}} \right), \tag{28}$$

$$\mathbf{m}^{k+1} = \mathbf{m}^{k} + \eta \left( \mathbf{m}_{wTV}^{k} - \mathbf{m}^{k} \right), \tag{29}$$

where $\eta \in [1, 2)$ controls convergence speed.

## IV. Experiments

### A. In vivo datasets

Comprehensive experiments are conducted to evaluate the performance of PAIR. Four datasets acquired on three vendors from four centers are employed for *in vivo* experiments. Their imaging parameters are shown in Table I.

For all datasets, odd-even EPI shifts have been corrected carefully. Coil sensitivity maps are calculated by ESPIRIT using $\mathbf{m_0}$ [40]. The directionally encoded color maps are produced using a Matlab toolbox[1].

### B. Experiments settings

For comparative study, navigator-based (IRIS [12]) and navigator-free (POCS-ICE [21], MUSSELS-IRLS-CS [17], PLRHM [18]) methods are adopted. POCS-ICE is an explicit phase reconstruction method. MUSSELS-IRLS-CS and PLRHM are implicit phase reconstruction methods which introduce the low-rankness priors into reconstruction. All the reconstructed multi-shot images by implicit phase methods are displayed after combining by the sum of squares. POCS-ICE and PLRHM are implemented by ourselves, and MUSSELS-IRLS-CS is provided by Dr. Mathews Jacob online[2]. All methods are implemented in MATLAB and the parameters are optimized for best performance in terms of least artifacts.

We take peak signal-to-noise (PSNR), and average angular error (AAE) [16] as objective criteria:

---

[1] https://www.mathworks.com/matlabcentral/fileexchange/34008-dti-fiber-tractography-streamline-tracking-technique
[2] https://github.com/sajanglingala/data_adaptive_recon_MRI

TABLE I IMAGING PARAMETERS OF THREE DATASETS.

| Dataset | Vendor/Center | Channel | Shot | Matrix Size | Resolution (mm³) | B-values (s/mm²) | Signal average | Diffusion directions |
|---------|---------------|---------|------|-------------|------------------|------------------|----------------|---------------------|
| I | Philips 3.0T /Beijing, China. | 8 | 8 | 230×232 | 1.0×1.0×4 | 0, 800 | 1 | 15 |
| II | UI 3.0T /Shanghai, China | 17 | 4/8 | 160×160, 230×224 | 1.4×1.4×5, 1.0×1.0×5 | 0, 1000, 2000, 3000, 4000 | 1 | 3 |
| III | Philips 3.0T /Xiamen, China | 32 | 4 | 180×180 | 1.2×1.2×5 | 0, 1000 | 2 | 12 |
| VI | XinGaoYi 1.5T /Yuyao, China | 16 | 4 | 140×192 | / | 0, 1000 | 2 | 3 |

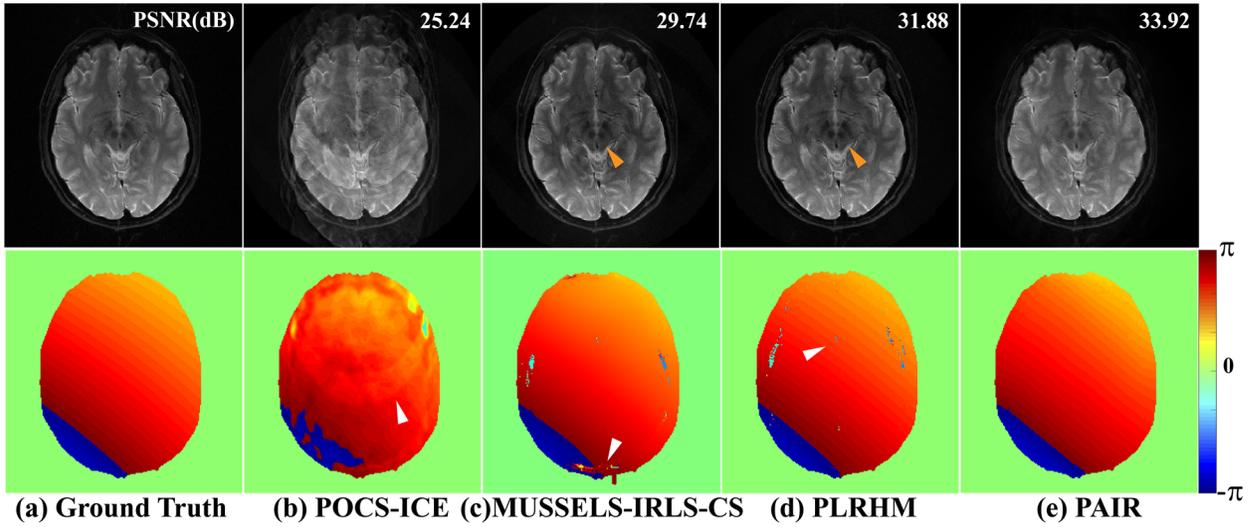

**(a) Ground Truth    (b) POCS-ICE  (c)MUSSELS-IRLS-CS   (d) PLRHM    (e) PAIR**

Fig. 7. Reconstructed phase and magnitude images of the simulated eight-shot eight-channel data. The top row are magnitude images and the bottom row are the representative shot phases. Phase and magnitude errors have been remarked with yellow and white arrows, respectively. PSNRs are in the upper right corner.

$$\text{PSNR (dB)} := 10 \cdot \log_{10}\left(\frac{NM}{\|\hat{\mathbf{m}} - \mathbf{m}\|_2^2}\right), \quad (30)$$

$$\text{AAE (degree)} := \frac{1}{L}\sum_{n=1}^{N} \cos^{-1}\left(\langle \mathbf{v}_l \cdot \hat{\mathbf{v}}_l \rangle\right) * 180/\pi, \quad (31)$$

where $\|\ \|_2$ is the $l_2$ norm, $\mathbf{m}$ and $\hat{\mathbf{m}}$ are vectorized reference and reconstructed images, respectively. $\mathbf{v}_l$ and $\hat{\mathbf{v}}_l$ represent the primary diffusion direction vector of reference and reconstructed maps. $L$ is the number of vectors. The higher PSNR and lower AAE indicate a lower noise level and smaller angular error, respectively.

### C. Shot phase estimation

We test the performance of PAIR on the simulated 8-shot 8-channel data (Fig. 7). The simulated process is as same as that in Fig. 5. The reconstructions in Fig. 7 are challenging because the number of shots is as high as the number of channels, which has not been reconstructed by navigator-free methods before.

POCS-ICE has some severe artifacts in the magnitude image and obvious phase aliasing (white arrows in Figs. 7(b)). Some slight dark shadows and phase errors exist in the magnitude and phase image of MUSSELS-IRLS-CS and PLRHM (yellow and white arrows in Fig. 7 (c) and (d)).

The magnitude and phase of PAIR are most consistent with ground truth and PAIR has the highest PSNR (Fig. 7(e)).

### D. High-resolution DWI

Reconstructed results of navigator-based method IRIS are employed as references (Figs. 8(a)), because the shot phase is accurately estimated by the navigator and the reconstructed images have no obvious artifact. Some obvious artifacts can be observed in the results of POCS-ICE and MUSSELS-IRLS-CS (yellow arrows in Figs. 8(b) and (c)). PLRHM and PAIR have a good tolerance to artifacts. PLRHM shows a relatively low SNR. Some edges have been blurred, such as the posterior horn of lateral ventricles (Fig. 8(d)). The result of the proposed PAIR achieves a lower noise level and more clear edges (Fig. 8(e)).

Directionally encoded color maps are estimated with fifteen diffusion directions. They are calculated by fractional anisotropy images times the primary diffusion direction. Some color mismatches with reference maps could be observed in Figs. 8(a)-(d). The result of PAIR is closest to the reference maps visually and achieves the lowest angular error AAE.

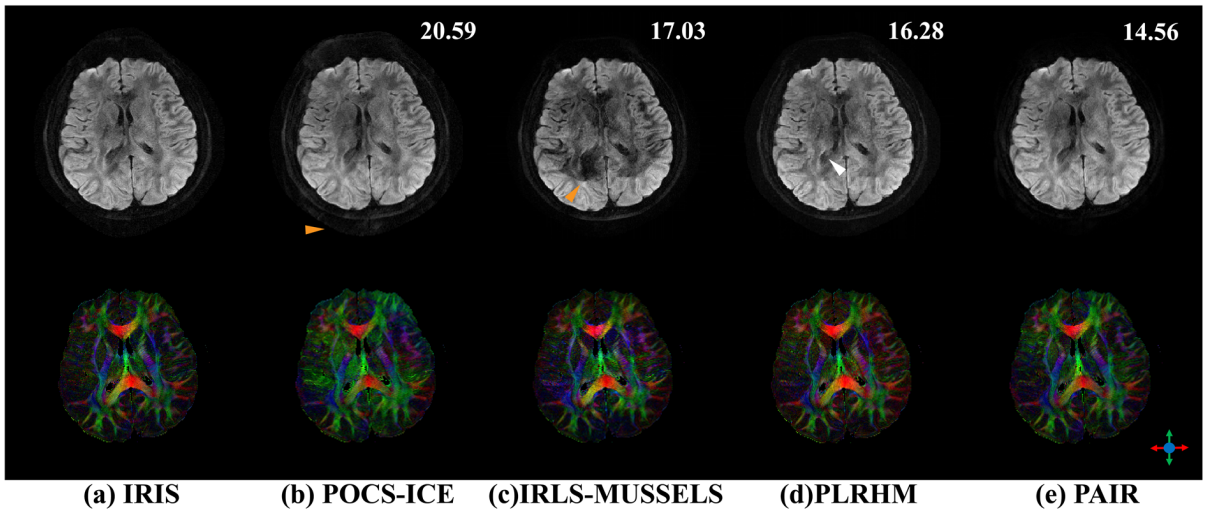

**Fig. 8.** Reconstructions of high-resolution DWI images and color fractional anisotropy images estimated from 15 diffusion directions in Dataset I. The top row are the first direction DWI and the bottom row are directionally encoded color maps. Artifacts and blurred edge have been remarked with yellow and white arrows, respectively. AAEs are in the upper right corner.

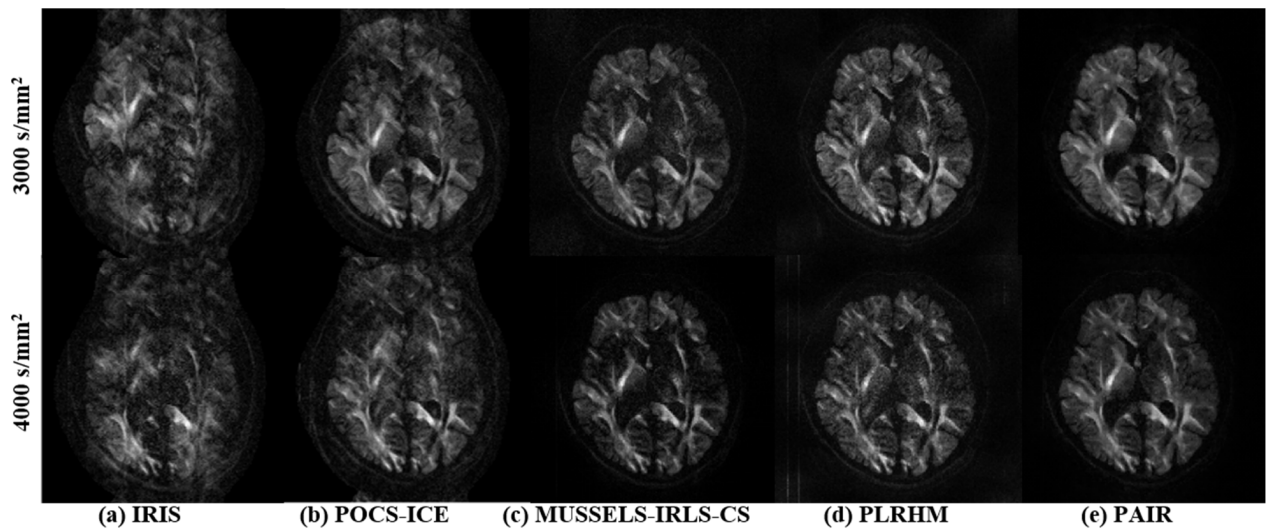

**Fig. 9.** Reconstructions of high b-values eight-shot data in Dataset II.

Thus, PAIR outperforms other state-of-the-art navigator-free methods on high-resolution DWI reconstruction.

### E. High b-values DWI

The ultra-high b-values DWI data (3000 and 4000 s/mm²) have a significantly low SNR, which poses a severe challenge for reconstruction. The navigator-based IRIS can hardly reconstruct the image (Fig. 9(a)), which may be caused by the low SNR and geometric mismatch between the navigator and image echo. POCS-ICE introduces obvious ghosting artifacts (Fig. 9(b)). Both MUSSELS-IRLS-CS and PLRHM better remove artifacts but suffer from relatively low SNR (Figs. 9(c) and (d)). The proposed PAIR outperforms other methods on much better tolerance to artifacts and noise (Fig. 9 (e)).

These results show that the PAIR is applicable for ultra-high b-values DWI reconstruction and does not need navigator signals.

### F. Accelerated DWI

For accelerated DWI with undersampling, the four-shot data in Dataset II is reconstructed with retrospectively uniform and partial Fourier undersampling. The sampling rate of them are 0.5 and 0.6, respectively. For evaluation, the fully sampled four-shot data are reconstructed as references (first row in Fig. 10).

On the fully sampled data, All the methods show comparable performances (first row in Fig. 10). Two explicit phase methods POCS-ICE and PAIR have relatively better resistance to zipper artifacts than MUSSELS-IRLS-CS and PLRHM.

On the undersampled data, error maps corresponding to the fully sampling references are calculated. Compared with other methods, PAIR provides the highest fidelity results on both uniform and partial Fourier undersampling reconstructions.

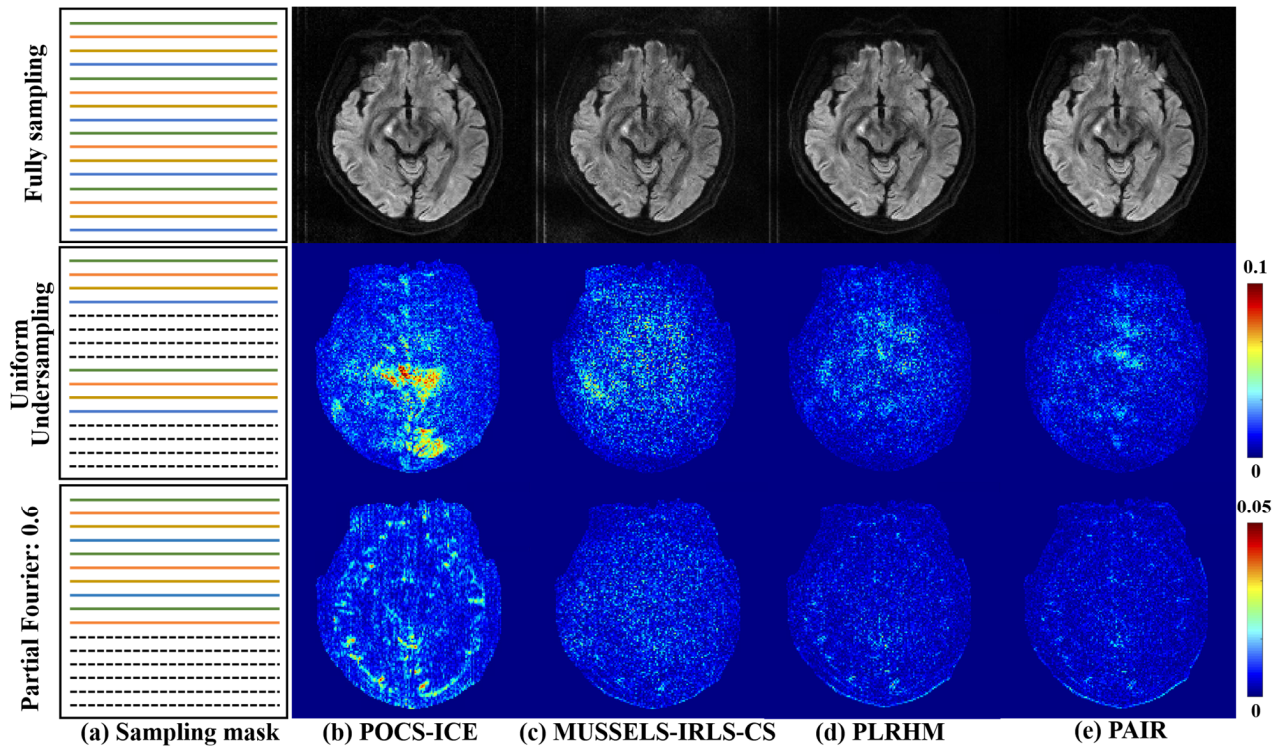

**Fig. 10.** Reconstructions of fully sampling and undersampling four-shot data in Dataset II. Note: Solid and dotted lines represent sampled shot and unsampled shot data, respectively. Different colors represent different shots.

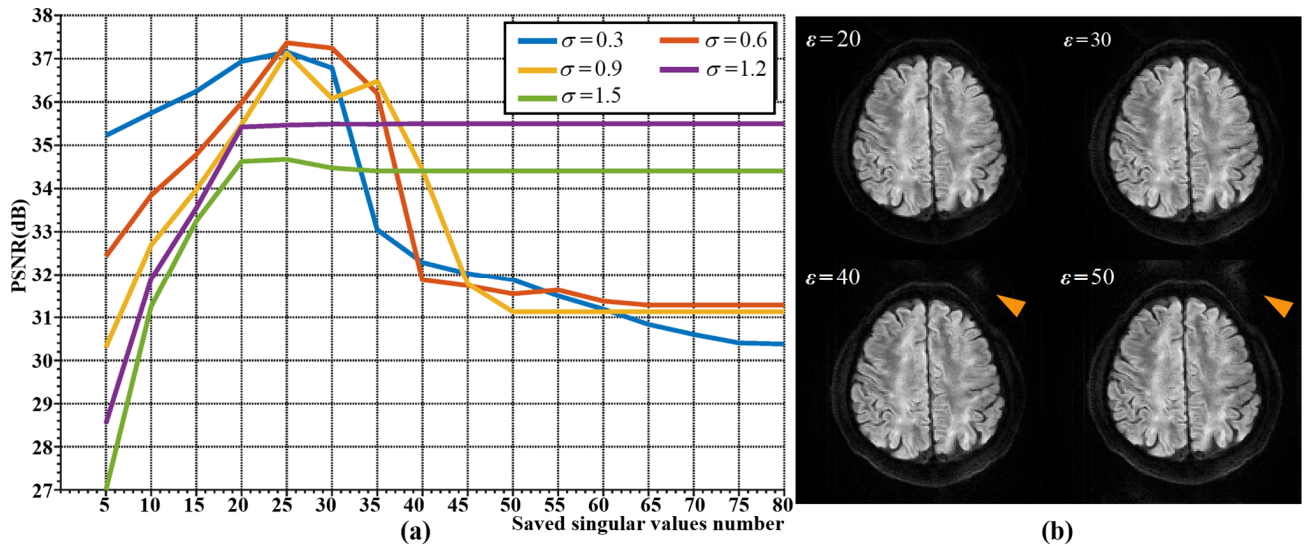

**Fig. 11.** PAIR Reconstructions on simulation and *in vivo* data with different $\varepsilon$ and $\sigma$. (a) The simulation data is 4-shot, 8-channel. (b) The *in vivo* data from Dataset II is 4-shot, 17-channel, 1.5×1.5 mm², 1000 s/mm², and $\sigma$ is 0.6.

## V. DISCUSSIONS

### A. Discussion on Parameter Settings

The effect of parameter settings in the singular value thresholding operator $\text{SVT}_\varepsilon(\mathbf{Z}, \sigma)$ is discussed in the simulation and *in vivo* experiments here. The operator is performed on the structured low-rank matrix $\mathbf{Z}$ to constrain phase smoothness. The first $\varepsilon$ singular values are saved and others minus a proper threshold $\sigma$.

In the simulation study, the radius of limited support $R$ is 2, and the size of $\mathbf{Z}$ is 104×98550. PAIR achieves nice PSNR when $\varepsilon$ is in the range of 20-30, under a proper $\sigma$ (0.3-0.9) (Fig. 11(a)). When $\sigma$ is too large ($\geq 1.2$), only singular values larger than $\sigma$ are saved, resulting in invariant PSNR (second half of the purple and green curve in Fig. 11(a)).

In *in vivo* study, the radius of limited support $R$ is 2, and the size of $\mathbf{Z}$ is 104×48050 and $\sigma$ is 0.6. In a wide range of $\varepsilon$ (10-30), PAIR has comparable results (Fig. 11(b)). When $\varepsilon$ becomes larger (40-60), some artifacts will remain in the images (yellow arrows in Fig. 11(b)).

The above experiments indicate PAIR is insensitive to parameters and has robust performance.

Other typical parameters are $\lambda = 1$, $\beta \in [10^{-4}, 10^{-1}]$, $\eta = 1.5$, the iteration stop condition is $\left\| \mathbf{m}^{k+1} - \mathbf{m}^{k} \right\|_{F}^{2} / \left\| \mathbf{m}^{k} \right\|_{F}^{2} \geq 10^{-5}$, and the max iteration is 1000.

### B. Discussion on weighted TV

The effect of weighted TV has been analyzed in the simulation study (Fig. 5). We compare PHASE and PAIR on *in vivo* data to study the role of the weighted TV further.

The high resolution (1.0×1.0 mm²) and high b-values (3000 s/mm²) show that, compared with PHASE, weighted TV provides PAIR better noise suppression while preserving edges.

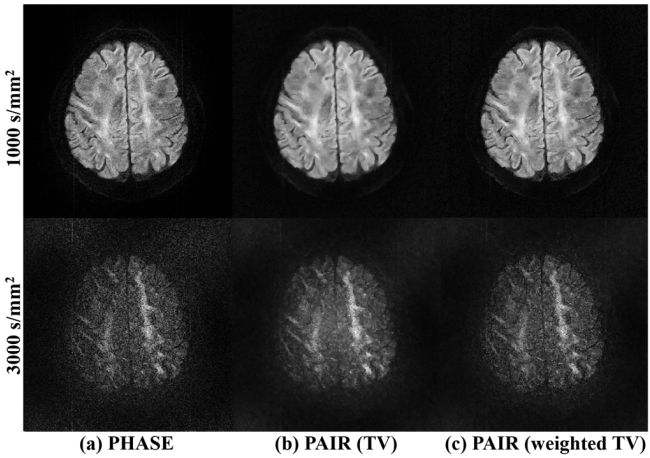

**(a) PHASE**　　**(b) PAIR (TV)**　　**(c) PAIR (weighted TV)**

Fig. 12. Reconstructions by PHASE, PAIR with TV, and PAIR with weighted TV. The data from Dataset II are 4-shot, 17-channel, 1.0×1.0 mm², 1000 and 3000 s/mm².

### C. Discussion on MUSSELS family

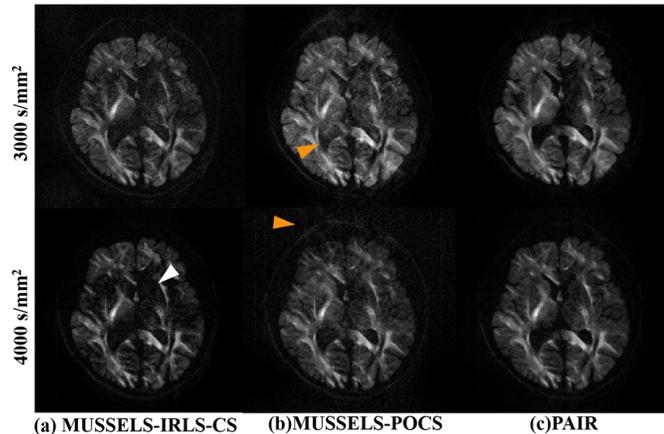

**(a) MUSSELS-IRLS-CS**　**(b)MUSSELS-POCS**　**(c)PAIR**

Fig. 13. Reconstructions of DWI data from Dataset II. The data are 4-shot, 17-channel, 1.5×1.5 mm², b-value 3000 and 4000 s/mm².

We compare PAIR with two kinds of MUSSELS variants to study the effect of solving algorithms. MUSSELS-IRLS-CS is a computationally efficient MUSSELS formulation with conjugate symmetry (CS) by modifying the iterative reweighted least squares (IRLS). MUSSELS-POCS is a MUSSELS formulation solved by the POCS algorithm.

The *in vivo* study shows that MUSSELS-IRLS-CS has slight signal loss (white arrows in Fig. 13), while some artifacts residual exist on MUSSELS-POCS reconstructions (yellow arrows in Fig. 13).

Compared with two MUSSELS variants, the proposed PAIR has better SNR and higher fidelity reconstruction.

### D. Discussion on multi-vendor multi-center reconstruction

Multi-center data harmonization is an important issue for healthcare studies [41, 42]. We test the performance of PAIR on diverse DWI data acquired by scanners from 3 vendors in 4 centers (Fig. 14).

PAIR shows robust performance on the above multi-vendor multi-center data, and the motion artifacts could be removed well (Fig. 14).

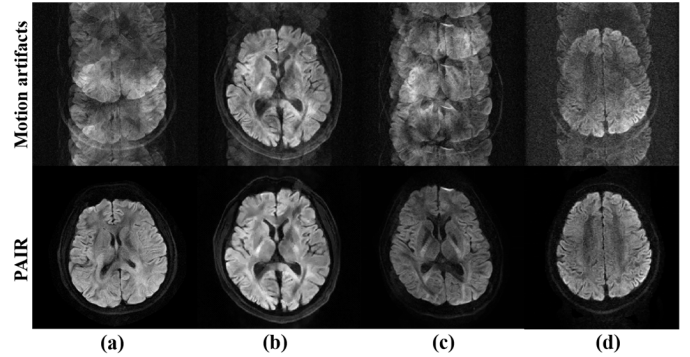

**(a)**　　**(b)**　　**(c)**　　**(d)**

Fig. 14. PAIR reconstructions on the multi-center multi-vendor in vivo data. (a) is from Dataset I: Philips 3.0T scanner in Beijing, China. 8-shot, 8-channel, b-value 800 s/mm², in-plane resolution 1.0×1.0 mm². (b) is from Dataset II: United Imaging 3.0T scanner in Shanghai, China. 4-shot, 17-channel, b-value 1000 s/mm², in-plane resolution 1.5×1.5 mm². (c) is from Dataset III: Philips 3.0T scanner in Xiamen, China. 4-shot, 32-channel, b-value 1000 s/mm², in-plane resolution 1.2×1.2 mm². (d) is from Dataset VI: XinGaoYi 1.5T scanner in Yuyao, China. 4-shot, 8-channel, b-value 1000 s/mm².

### E. Discussion on reconstruction time of PAIR

The reconstruction time of PAIR is tested on the three kinds of DWI data (Tab. II). The iteration stop condition is $\left\| \mathbf{m}^{k+1} - \mathbf{m}^{k} \right\|_{F}^{2} / \left\| \mathbf{m}^{k} \right\|_{F}^{2} \geq 10^{-5}$, and the max iterations is 1000. On all three data, PAIR could reach convergence and complete reconstruction quickly (Tab. II).

All the computation procedures are executed by MATLAB, running on a CentOS 7 computation server with twenty Intel Core i9-9900X CPUs of 3.5 GHz and 125 GB RAM. No parallel computation is employed.

TABLE II. **RECONSTRUCTION TIME OF PAIR**

| Dataset | Matrix size (RO×PE×Channel×Shot) | Iterations when stop | Time (s) |
|---|---|---|---|
| II | 160×160×17×4 | 22 | 12.4 |
| II | 230×224×17×4 | 27 | 26.1 |
| III | 180×180×32×4 | 45 | 34.3 |

## VI. Conclusion and Outlook

In this work, we aim at solving the challenging multi-shot DWI reconstructions under inter-shot motions between shots and a low signal-to-noise ratio. A model with paired phase and magnitude priors is proposed to regularize the reconstruction. The former prior is derived from the smoothness of the shot phase and enforced with low-rankness in the k-space domain. The latter explores similar edges among multi-b-value and multi-direction DWI with weighted total variation in the image domain. Comprehensive experiments show that PAIR has stable shot phase estimation and reconstruction performance on high shot number data (8 shots). Compared with state-of-the-art methods, it shows much better and more robust performance on some low SNR scenarios, such as the undersampling (uniform and partial Fourier) and ultra-high b-values DWI (4000 s/mm$^2$). Moreover, reconstructions on multi-vendor multi-center DWI data indicate its nice robustness.

The good performance and nice robustness of PAIR on high-resolution, ultra-high b-value and accelerated DWI reconstructions show great potential for advanced clinical DWI applications and brain function research.

In the further, we will develop PAIR to solve abdomen DWI reconstructions, such as high-resolution prostate and liver DWI. These reconstructions suffer from lower SNR and non-smooth shot phase, which may be challenging for PAIR and need great improvements.

In addition, to make it easier to use PAIR, we have implemented and deployed PAIR on the open-access cloud platform, CloudBrain-ReconAI [43-46] (please visit https://csrc.xmu.edu.cn/ReconAndEval.html).

### Acknowledgements

The authors thank Yiman Huang, Dicheng Chen for the discussions on the MRI reconstruction; Dr. Hua Guo, Yi Xiao for their assistances in data acquisition and image analysis; Yu Hu for creating video tutorials on CloudBrain-ReconAI usage; Dr. Mathews Jacob for sharing the IRLS-MUSSELS code online; The authors also thank the editors and reviewers for the valuable suggestions.